# A Single-Level Tunnel Model to Account for Electrical Transport through Single Molecule- and Self-Assembled Monolayer-based Junctions

*Alvar R. Garrigues[1], Li Yuan[2], Lejia Wang[2], Eduardo R. Mucciolo[1], Damien Thompson[3,4], Enrique del Barco[1]\* & Christian A. Nijhuis[2,5]\**

[1] Department of Physics, University of Central Florida, Orlando, Florida 32816 - USA

[2] Department of Chemistry, National University of Singapore, 3 Science Drive 3, Singapore 117543

[3] Department of Physics and Energy, University of Limerick, Ireland

[4] Materials and Surface Science Institute, University of Limerick, Ireland

[5] Centre for Advanced 2D Materials, National University of Singapore, 6 Science Drive 2, Singapore 117546

(\*) CORRESPONDING AUTHORS: E.d.B: Tel.: (+1) 823 0755, e-mail: delbarco@ucf.edu; C.A.N: Tel.: (+65) 6516 2667, e-mail: christian.nijhuis@nus.edu.sg

KEYWORDS: Molecular tunnel junctions, single-electron transistors, molecular electronics, molecular diodes, molecular rectification, electron transport.

**Abstract:**

We present a theoretical analysis aimed at understanding electrical conduction in molecular tunnel junctions. We focus on discussing the validity of coherent versus incoherent theoretical formulations for single-level tunneling to explain experimental results obtained under a wide range of experimental conditions, including measurements in individual molecules connecting the leads of electromigrated single-electron transistors and junctions of self-assembled monolayers (SAM) of molecules sandwiched between two macroscopic contacts. We show that the restriction of transport through a single level in solid state junctions (no solvent) makes coherent and incoherent tunneling formalisms indistinguishable when only one level participates in transport. Similar to Marcus relaxation processes in wet electrochemistry, the thermal broadening of the Fermi distribution describing the electronic occupation energies in the electrodes accounts for the exponential dependence of the tunneling current on temperature. We demonstrate that a single-level tunnel model satisfactorily explains experimental results obtained in three different molecular junctions (both single-molecule and SAM-based) formed by ferrocene-based molecules. Among other things, we use the model to map the electrostatic potential profile in EGaIn-based SAM junctions in which the ferrocene unit is placed at different positions within the molecule, and we find that electrical screening gives rise to a strongly non-linear profile across the junction.

**INTRODUCTION**

The last decade has led to substantial advances in the understanding of electrical conduction through molecular systems, mostly due to an improved control of the different measuring techniques employed to study molecular junctions, including electromigrated single-electron transistors (SETs) [1-8], scanning tunneling microscopy (STM) break-junctions [9-10], and techniques enabling measuring self-assembled monolayers (SAMs) of molecules sandwiched between two macroscopic electrodes [11-20]. By far most fabrication techniques produce two-terminal junctions in the form of electrode-molecule-electrode, with either a single molecule or a self-assembled monolayer (SAM) as the active component. However, the lack of an electrical gating capability in two-terminal junctions complicates determination of the energy level alignment. STM break-junctions can be used to obtain statistics on electron transport at an individual molecular level but the supramolecular and the electronic structure of the junctions cannot be investigated independently. By contrast, it is possible to measure both the supramolecular and electronic structure of the SAMs immobilized on one of the electrodes (usually referred to as the bottom-electrode) using standard surface characterization techniques before fabrication of the top-contact. For most systems, however, it is not clear how much the energy level alignment changes once the molecules are in contact with the top-electrodes [21] and therefore models are often used to extract the relevant transport parameters. From a technological point of view, SAM-based junctions are promising candidates for molecular electronics technologies, with rectification ratios rapidly approaching values of commercial semiconductor devices [22,23]. On the other hand, three-terminal SETs electrically gate the molecular electronic states of an individual molecule, enabling the full spectroscopic resolution of the molecular energy landscape [1-8]. Comparative studies using these different techniques are required to establish the principles governing electrical conduction through molecular tunnel junctions under a wide range of experimental conditions.

Despite the increasing number of experimental reports, and to a lesser extent theoretical works, dedicated to the topic of electrical conduction through molecular junctions (see Refs.[24-26] and references therein), there remains the challenge of extracting fundamental characteristic junction parameters from the experiments in a consistent way. The complexity of the real devices, with large variations in the parameters that affect electrical conduction, and the different possible transport regimes, make the rigorous treatment of electron transport through junctions extremely difficult. In this context, first-principles calculations have the advantage that one can account for structural details of the junctions [27-28]. However, they are not generally applicable (*i.e.*, they need to be tailored to the specific junction under study) and may require very long calculation times even on high-performance computing platforms. Alternatively, analytical models dealing mainly with the energetics of the junction, such as the Landauer formalism for coherent tunneling (or the rate equations for incoherent tunneling), are easy to employ but rely on rough approximations of some critical parameters and may miss some of the physics involved in the process, particularly when conduction is driven or assisted by interactions with the molecular surroundings (*e.g.*, reorganization energies, many-body interactions, energy level renormalization, image charges, *etc*.).

There are ongoing theoretical efforts focused on solving the problem by integrating different effects to formulate a non-equilibrium theory in terms of molecular states that includes many-body correlations when computing electron transport through molecular junctions (for a detailed discussion on this see Ref. [26]). This is far from the goal of the present work, where we do not attempt to develop a new theory but instead aim to test the applicability of a well-known single-level tunneling model and variations thereof [29-32], to capture the essential physics of molecular junctions. The key metric for success of our model is the ability to easily implement it using *experimentally* obtained input parameters. In this article, we test this single-level tunneling model using properties of molecular diodes measured in two different test-beds: *i*) Three-terminal SET experiments performed at different temperatures on individual $S(CH_2)_4Fc(CH_2)_4S$ molecules; and *ii*) experiments in SAM-based EGaIn junctions of $S(CH_2)_{11}Fc_2$ molecules where two distinct molecular levels participate in charge transport [22]. In addition, we employ the model to interpret experiments performed on SAM-based EGaIn junctions of $S(CH_2)_nFcC_{13-n}$ molecules where the ferrocene (Fc) unit is placed at different positions (n) within the alkyl chain [33]. Fitting our model to the experimental data enables the determination of the electrostatic potential profile in the junction, which we find to be strongly non-linear, indicating the presence of significant electrical screening in the junction. Comparison between the different junctions shows that the single-level tunneling model satisfactorily describes both single molecule and SAM based junctions. The model captures the essential physics, including the asymmetry of the *J*(V) curves, the shape of the electrostatic potential profile and the temperature dependence of the transport excitations. Finally, we use this model to discuss in detail the range of validity of coherent versus incoherent theoretical formulations that have been used to rationalise some recent experimental results.

**RESULTS**

**The single-level transport model: Coherent versus incoherent tunneling**

In the simplest approximation, a molecular junction can be represented by the schematic diagrams shown in Figure 1, where the molecule sandwiched between two electrodes (three in the case of a SET) is represented by a discrete set of levels (with N representing the number of electrons in each level) separated from the electrostatic potential in the electrodes by tunnel barriers. The molecular levels closest to the electrostatic potentials of the leads at zero bias are known as the HOMO (highest occupied molecular orbital, N) and LUMO (lowest unoccupied molecular orbital, N+1) levels and are well separated in energy by a ~2-5 eV conductance gap in solid state junctions. The level closest to the Fermi energy of the electrodes governs the electrical conduction through the junctions for moderate bias potentials.

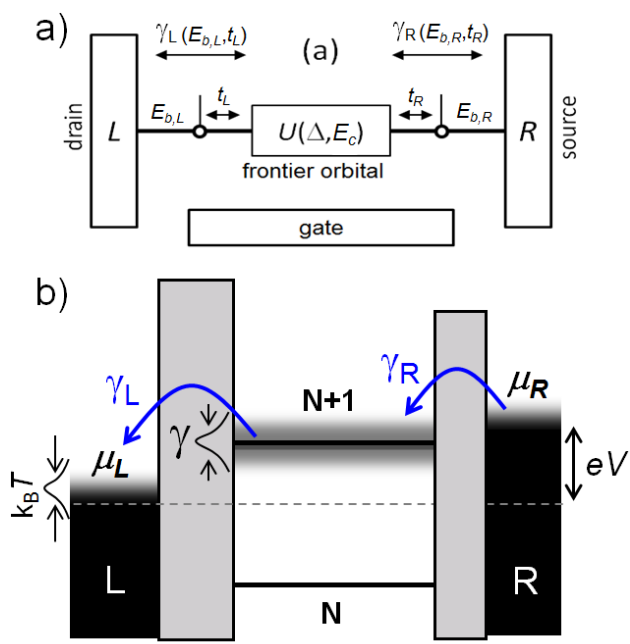

**Figure 1: A molecular tunnel junction**. Schematics of the electrical coupling (**a**) and corresponding energy level alignment (**b**) in a molecular junction composed of a molecule with multiple electrostatic levels sandwiched in between two electrodes (L (left): drain and R (right): source), including all the functional parameters that govern electron tunneling through the junction. These are the tunneling rates, $\gamma_{L/R}$, the binding energies, $E_{b,L/R}$, and the intermolecular couplings, $t_{L,R}$, between the molecule and left and right electrodes, with $\gamma = \gamma_L + \gamma_R$ being the level width. In addition, $\Delta$ represents the intramolecular level spacing, $E_c$ the charging energy, and $\mu_{L/R}$ the electrostatic potentials of the leads.

Conduction through a molecule can, in principle, be understood in terms of sequential tunneling (either coherent or incoherent) involving one molecular level. In such a process the electron tunnels from the source into the molecule at a rate $\gamma_R$ and then out into the drain lead at a rate $\gamma_L$. The ratio between these rates and the characteristic intramolecular relaxation times determines whether the electron transport occurs coherently or incoherently. This means that if the electron spends sufficient time inside the molecule to significantly couple with internal/external degrees of freedom (*i.e.*, short relaxation time and short decoherence time relative to the tunneling rates) then the process will become incoherent, with an associated loss of phase in the case of elastic tunneling, and loss of both phase and energy in the case of inelastic tunneling. In wet electrochemical molecular charge-transfer processes, energy relaxation is usually facilitated by intramolecular vibrational modes and polaron excitations in the surrounding solvent molecules. In terms of Marcus theory [34], these are known as inner- and outer-sphere molecular reorganization processes, respectively, and are characterized by an activation energy arising from the classical energy barrier governing the process. At sufficiently high temperatures, these processes lead to an exponential increase of the charge transfer (accompanied by an Arrhenius law and incoherent tunneling). However, in the junctions discussed in this work, where the molecules are present in solid-state devices and charge transport is studied in the absence of solvent, outer-sphere reorganization processes do not likely play an important role in the electrical conduction through the junction. This is not necessarily the case in all SAM-based junctions, since even in the absence of solvent molecules, neighboring molecules can play this role, particularly when considering complex molecules with high degrees of polarizability.

We expect that inner-sphere reorganization will always be important, and can be related to intra-molecular vibrational modes with energies usually below a few tens of millielectronvolts (<30 meV), which is smaller than or comparable to other energy scales relevant to the conduction process, such as thermal broadening at high temperatures ($k_BT$ = 13-86 meV in the range 100-1000K). Thus, when only one level participates in the conduction through the junction (which may be the case in many molecular junctions), the vibrational modes can be formally incorporated into an effective temperature-dependent level broadening (the higher the temperature the more vibrational modes will be sampled), which means that it is reasonable to still treat the transport process with simple single-level transport models, as we do in the present work. Note the different causes of level broadening, which we can separate into two main contributions: i) broadening due to the coupling of the molecule to the electrodes, with tunneling rate $\gamma = \gamma_L + \gamma_R$, and, ii) broadening arising from the coupling of the electron to other degrees of freedom of the molecule, $\gamma_0$. Of course, this is just a simplification (virtual transitions to high-energy molecular states take place during the conduction process, leading to renormalized parameters) and should be taken as such. However, this picture can be justified from a more fundamental calculation involving a non-equilibrium quantum mechanical treatment of the response of the molecule to its coupling to leads. When a single energy level is active in the molecule, and the coherent interactions between electrons in the molecule and those in the leads are negligible, the non-equilibrium expression for the current reduces to an expression similar to the Landauer-Büttiker formula [30], namely,

$$I = \frac{2q}{h} \int dE \left[ f\left(\frac{E-\mu_L}{k_BT}\right) - f\left(\frac{E-\mu_R}{k_BT}\right) \right] Tr\left\{ \frac{\gamma_L(E)\gamma_R(E)}{\gamma_L(E)+\gamma_R(E)} Im[G^r(E)] \right\}, \qquad (1)$$

where $f(x) = 1/(e^x + 1)$, $\mu_L$ and $\mu_R$ are the chemical potential at the leads, $\gamma_L(E)$ and $\gamma_R(E)$ are the partial level widths due to the coupling to right and left leads, respectively, and $G^r$ is the molecule's single-particle retarded Green's function, which contains a self-energy that takes into account the coupling to leads and to other degrees of freedom as an imaginary part, namely, a total level width. In fact, this has been the starting point of several successful theoretical descriptions of molecular electronic transport experiments (see, *e.g.*, Refs. [35] and [36]).

We acknowledge that there may be other processes playing important roles in the transport through solid-state junctions, such as intermolecular interactions or electrostatic potential changes due to electron charging or image-charge effects [27], among others. Some of these could be incorporated into simple single-level transport models. As we show below, for a single-level junction system the coherent and incoherent tunneling treatments provide identical predictions of the current at all temperatures, provided that one incorporates into the level broadening the two effects discussed above concerning outer-sphere reorganization and vibrational modes. Therefore, even a single-level coherent transport model can explain the exponential thermal enhancement of the conductance through a molecular junction without the need to invoke Marcus relaxation processes. This is because the temperature dependence of the conductance naturally arises from the thermal broadening of the energy of the electrons in the leads (see, *e.g.*, the work by van der Zant *et al.* [37]). In other words, a transition between a plateau-like to temperature-dependent conductance in a molecular junction should not be taken as conclusive proof of inelastic incoherent tunneling. Sequential tunneling may well be coherent and one simply cannot distinguish them when transport involves a single energy level without considering the ratio of the relaxation rates involved.

As mentioned before, the transition between coherent and incoherent tunneling is determined by the tunneling rates in and out of the molecule (when compared with the intrinsic molecular relaxation rates). As explained by Moth-Poulsen and Bjørnholm in their review article [38], the tunneling rates for a complex molecule usually depend on the strength of the bond of the molecule to the electrodes, $E_{b,i}$, and on the intramolecular coupling, $t_i$, between the electrode and the molecular frontier orbital (see *Fig. 1a*), and are, in general, difficult to engineer. As mentioned above, the molecule-electrode couplings result in broadening of the molecular levels ($\gamma = \gamma_L + \gamma_R$). The capacitive couplings of the frontier orbital with the electrodes determine the electric potential drops at both sides of the active conduction unit ($V_L$ and $V_R$) and provide asymmetry to the junctions when $L_L \neq L_R$. This asymmetry is parameterized by the dimensionless division parameter $\eta = V_R/(V_L + V_R)$, which gives the ratio of the voltage drop between the molecule and the right electrode with respect to the total voltage drop in the junction.

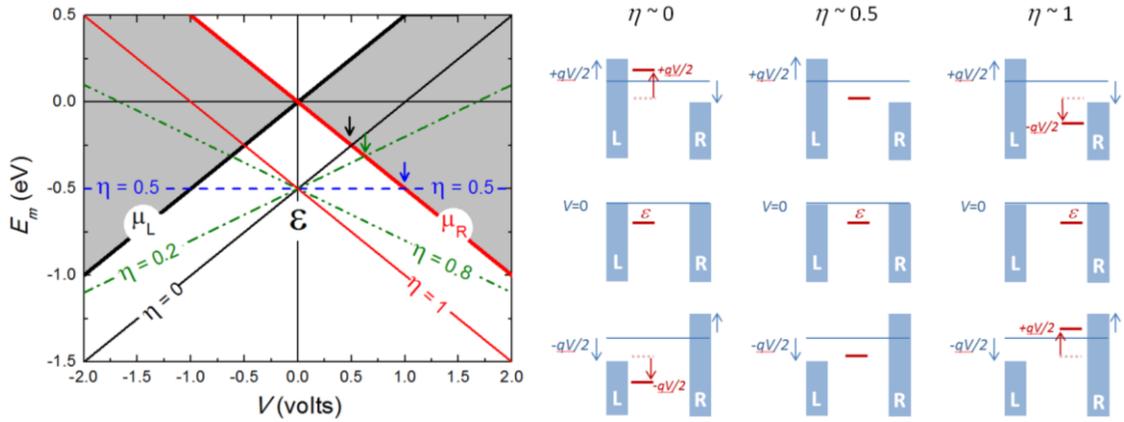

**Figure 2: Electrostatic coupling in a tunnel junction**. (**a**) Voltage dependence of the energy of the molecular frontier orbital $E_m$ for different values of the voltage division parameter $\eta$, with respect to the electrochemical potentials $\mu_L$ and $\mu_R$ of the electrodes in the junction. $\varepsilon$ is the zero-bias energy offset between the molecular orbital and the electrodes. The arrows show when the molecular level enters the conduction window (grey areas). (**b**) Represent the energetic configurations of the molecular level and the electrochemical potential of the electrodes for three representative values of the voltage division parameter, *i.e.*, $\eta = 0$, 0.5 and 1.

The molecular frontier orbital follows the average effective potential within the molecule and its energy can be expressed as $E_m = \mu_R(V) + \varepsilon + \eta qV$, where $\varepsilon$ is the zero-bias energy offset between the molecular orbital and the electrodes, q is the electron charge, and $\mu_R$ is the electrochemical potential of the right electrode. The behaviour of $E_m$, $\mu_L$ and $\mu_R$ with respect to the bias potential $V$ is shown in Figure 2 for different values of $\eta$. For $\eta \sim 0.5$ the energy of the level remains constant as the bias potential is increased, while for $\eta \sim 0$ or $\sim 1$, the molecular level closely follows the right or left potentials, respectively.

Finally, thermal broadening of the energy of the electrons in the leads also greatly affects conduction through the junction and needs to be taken into account. When a bias potential is applied, two distinct Fermi distributions need to be used, one for each lead:

$$f_{L,R}(E) = \frac{1}{1+\exp\left[(E-\mu_{L,R})/k_B T\right]} \qquad (2)$$

In the case of a single conduction level, the electrical current can be calculated by solving the rate equations for each side of the molecule (see Supplemental Information for a generalized treatment of the rate equations for a molecule with multiple levels). In the limit of very small broadening ($\gamma \ll k_B T, |\varepsilon|, |\mu_L - \mu_R|$) one obtains the following electrical current expression (see detailed discussion in Supplemental Information and in Refs. [24,39]):

$$I_{L,R}(E) = (-q)\frac{\gamma_{L,R}}{\hbar}(f_{L,R}(E) - N), \qquad (3)$$

where $N$ is the occupation number or average number of electrons in the steady state. By definition, in this state there is no net transfer of electrons in or out of the junction (*i.e.*, $I_L + I_R = 0$), from which one can extract an expression for N:

$$N = \frac{\gamma_L f_L(E) + \gamma_R f_R(E)}{\gamma_L + \gamma_R} \qquad (4)$$

Substituting this functional for $N$ in Eqn. (3) one obtains the steady-state current through the junction (without taking into account the spin) as

$$I = I_L = -I_R = \frac{2q}{\hbar}\frac{\gamma_L \gamma_R}{\gamma_L + \gamma_R}[f_L(E) - f_R(E)]. \qquad (5)$$

Note that this treatment accounts for sequential tunneling of the electron from one of the leads and into the molecule and further into the opposite lead and is incoherent in nature. In other words, the rate equations formulation assumes that the electron's phase memory is completely lost once it tunnels into the molecule. That can only happen through relaxation processes — purely elastic in the case of a single level (no other levels to transit into), but possibly even inelastic. Apart from this, the first conclusion that one can extract from this simple expression is that current will only flow through the junction if a molecular level lies between the electrochemical potentials of the left and right electrodes (*i.e.*, within the conductance window). Of course, this approximation fails when the level width surpasses such energy separation, since it will lie partially out of the conductance window. In other words, the current will not increase indefinitely by simply increasing the level width (*i.e.*, increasing the coupling to the electrodes). On the other hand, the resistance has a quantum mechanical limit, which for a single transport channel translates into a universal quantum of conductance, $G_0$, as originally predicted by Landauer [40].

Interestingly, it is possible to relax the condition of very small broadening $\gamma \ll k_B T, |\varepsilon|, |\mu_L - \mu_R|$ and incorporate the effect of a finite level broadening into the rate equation calculation by introducing a broadened density of states (DOS) in the shape of a Lorentzian centered at the energy level $\varepsilon$,

$$D_\varepsilon(E) = \frac{\gamma/2\pi}{(E-\varepsilon)^2 + (\gamma/2)^2}. \tag{6}$$

The result is

$$I = \frac{q}{h} \int_{-\infty}^{\infty} dE\, D_\varepsilon(E) \frac{\gamma_L \gamma_R}{\gamma_L + \gamma_R} [f_L(E) - f_R(E)], \tag{7}$$

This expression matches exactly the one derived by Jauho, Wingreen, and Meir in 1994 [30] using a fully coherent formulation based on the Keldysh Green's function formalism (see Supplemental Information for a detailed discussion of this point). This may seem remarkable, but we note that for a single channel and a single level in the molecule, interference plays no role in the electron conduction through the molecule. As a result, coherent and incoherent sequential tunneling cannot be distinguished at a formal level. Both models provide the same level of accuracy in describing transport at low bias through a single level, as was verified in a pioneering experiment by the Saclay group [41]. In the standard derivation of Eqn. (7) in the context of coherent tunneling, the level broadening is entirely given by the sum of the partial broadenings due to leakage through the leads ($\gamma = \gamma_R + \gamma_L$). In the derivation based on rate equations (incoherent sequential tunneling) one can introduce an additional source of broadening unrelated to leakage, namely, $\gamma = \gamma_R + \gamma_L + \gamma_0$. Other than that, the two results are formally identical at any temperature or bias voltage [42].

Landauer derived an equation similar to Eqn. (7) under the assumption that the entire system (leads and "molecule") were one dimensional. But now we know that the validity of Eqn. (7) does not depend on the spatial dimensions, but rather on the number of channels involved. Namely, it is a correct description only in the case of a single channel in and out of the molecule on each lead. The derivation presented here is also only valid in the case of spinless charge carriers. It is straightforward to extend it to include spin ½ carriers; provided that coherence does not extend beyond the molecule, both rate equation and fully-coherent formulations yield again identical results. When the charging energy is strong and forbids double occupancy of the molecular level (that is, $E_c \gg k_B T, |\mu_L - \mu_R|$, which is typically the case for solid-state molecular junctions), the expression for the current becomes

$$I = \frac{2q}{h} \int_{-\infty}^{\infty} dE\, D_\varepsilon(E) \frac{\gamma_L \gamma_R}{\gamma_L + \gamma_R + \Gamma_{0\to 1}} [f_L(E) - f_R(E)], \qquad (8)$$

Where the level width associated to the level lifetime is given by

$$\Gamma_{0\to 1} = \int_{-\infty}^{\infty} dE\, D_\varepsilon(E) [\gamma_L f_L(E) + \gamma_R f_R(E)]. \qquad (9)$$

In the linear regime and when the molecular level falls within the bias voltage window, $\Gamma_{0\to1} \approx \gamma_R + \gamma_L$ and one finds that the current matches closely that of Eqn. (7), with the prefactor of 2 cancelling out. Notice that, in the opposite limit, when charging energy is weak and double occupancy is permitted, one can simply adopt Eqn. (7) and account for spin degeneracy in the conduction process by multiplying the right-hand side by a factor of 2 (see Supplemental Information for details).

Thus, it is our claim that Eqn. (7) can be formally taken as a good approximation to describe conduction through a molecular junction at all temperatures provided that: i) Electrical conduction through the junction is mainly governed by a single level (no transitions to high-energy states); and, ii) that intra- and inter-molecular interactions can be incorporated into the level width (elastic processes). These conditions are likely to be applicable to solid state molecular junctions, where interaction with solvent molecules is not possible and Marcus-like energy relaxation processes are unlikely.

It is easy to see from Eqn. (7) that at low temperature ($f_L(E) - f_R(E) = 1$ if $\mu_L < E < \mu_R$, $= 0$ otherwise) and low bias ($\mu_L \sim \mu_R$), the maximum conductance (occurring when the energy level coincides with the average electrochemical potential, $\varepsilon = \mu$) will be

$$G_{max} = \frac{I}{V} = \frac{q^2}{h} \frac{4\gamma_L\gamma_R}{(\gamma_L+\gamma_R)^2} \qquad (10)$$

This expression coincides with the universal conductance $G_0 = \frac{q^2}{h}$, if the two rates are equal ($\gamma_L = \gamma_R$) [for the spinless case].

As a side note, one can break up the Landauer conductance into two contributions, since

$$G^{-1} = \frac{h}{q^2 M} \frac{1}{T} = \frac{h}{q^2 M} + \frac{h}{q^2 M} \frac{1-T}{T} = G_C^{-1} + G_T^{-1}, \qquad (11)$$

with M representing the number of channels and the characteristics of the junction (see Eqn. (8) when $\gamma_L \neq \gamma_R$). This is convenient in order to separate the contact resistance ($G_C^{-1}$) from the tunneling resistance ($G_T^{-1}$) in a molecular junction, and particularly relevant for AC impedance transport measurements, where one can distinguish experimentally between capacitive and resistive contributions in the transport through the junction [43].

An interesting exercise within the context of the discussion in this article is to solve the asymptotic limit of Eqn. (7) at high temperatures. To be more explicit, this pertains to the case when $\gamma \ll |E_m| \ll k_B T \ll qV$ (see Supplemental Information for a detailed derivation). In this case, Eqn. (7) reduces to:

$$I = \frac{2q}{\hbar} \frac{\gamma_L \gamma_R}{\gamma_L + \gamma_R} \left(\frac{qV}{k_B T}\right) e^{-E_m/k_B T}. \tag{12}$$

The current in this case shows an activation behavior, with the "activation energy" $\varepsilon$ being the offset between the energy level in the molecule and the Fermi energy in the leads. Note that a similar behaviour is expected from classical Marcus charge transfer processes, for which the activation energy has a completely different origin, as described above. We note that the activation energies arising from these two very different mechanisms may display different dependencies on the electric potentials applied to the junction, which therefore become powerful diagnostic tools to discern charge transport mechanisms. From Eqn. (12) it follows that the activation energy $\varepsilon$ would decrease with increasing bias voltage, as the electrostatic potential of the lead approaches the molecular level. This does not need to be the case in, *e.g.*, polaron-assisted conduction, where polarization of neighbouring molecules may be affected by the applied bias in intricate ways, resulting from its direct effect on the polaron dynamics.

Figure 3 shows tunneling calculations using Eqns. (5) and (7), to solve for the electrical current through a single molecular level described with the following characteristic parameters:

$\varepsilon = 0.4$ eV, $\eta = 0.5$, $V_G = 0$, and $\gamma_L = \gamma_R = 1$ meV (in Fig. 3a-c) and 10 meV (in Fig. 3d) and with different bias voltages: $V \in 0.4 - 0.7$ V. As can be clearly seen, there are two distinct regimes: decaying and saturated currents. The temperature of the transition between the two regimes (namely, the inflexion point where the plateau starts to develop) depends on the bias voltage: the lower the bias the higher the transition temperature (compare Fig. 3a-c). The same correspondence is observable for the slope in the thermally activated regime. As determined from Eqn. (12), the "activation energy" decreases with increasing bias (since the separation $\varepsilon$ between the molecular level and the electrochemical potential of the leads decreases with increasing bias). In addition, one can observe that increasing the level width smears out the difference between the two regimes (compare Figs. 3b and 3d). Important for our discussion is the fact that one can explain temperature-dependent electrical conductance in a molecular junction with the same functionality (exponential) and within the common temperature range of existing experiments without the need for a thermal activation derived from Marcus-like energy relaxation processes in the molecules (also possible but less likely in many solid state junctions). Indeed, this approach was successfully employed to explain the temperature dependence of the conductance through an individual sulfur end-functionalized tercyclohexylidene molecule in a SET device by van der Zant and collaborators in 2006 [37]. In that case, the width of the molecular level had to be adjusted for different bias voltages, from a few meV at low bias up to 30 meV as the system was brought into resonance at large bias. As discussed above, this effect can be understood in terms of a broadening of the molecular level width as a result of internal molecular vibrations excited by the increasing current when the system is brought closer to resonance.

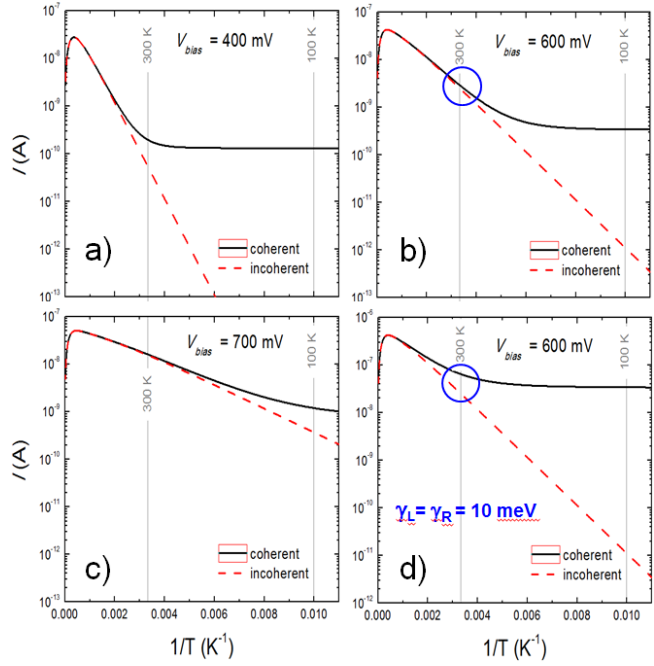

**Figure 3: Coherent vs. Incoherent tunnelling**. The electrical current through a single-level molecular junction as expressed by Eqns. (7) (broadened molecular level) and (5) (zero-width molecular level). The parameters used in the calculations are $\varepsilon = 0.4$ eV, $\eta = 0.5$, $V_G = 0$, and $\gamma_L = \gamma_R = 1$ meV for (**a-c**) and $\gamma_L = \gamma_R = 10$ meV for (**d**), which are typical values in the molecular junctions studied in this work. The plateau at low temperatures is a direct consequence of the broadening of the molecular level as given by Eqn.(6), which sets a maximum for a conductance at low temperatures.

**Three-terminal single-electron transistor measurements of a Ferrocene-based junction**

Figure 4a shows the electrical current at $T = 80$K as a function of gate voltage for a bias of $V = 10$ mV through an individual S-$(CH_2)_4$-Fc-$(CH_2)_4$-S molecule placed in between the leads of an electromigrated three-terminal $Al_2O_3$/Au SET (see sketch in Fig. 4) fabricated according to the procedure given in Ref. [44]. Two peaks in the current are clearly visible at $V = $ -0.3 V and +1.7 V. These peaks separate three consecutive charge states ($N$-1, $N$ and $N$+1) of the molecule, when no current is allowed to flow at low bias (*i.e.*, Coulomb blockade). Details of the transport results and the synthesis of the molecule are given in Ref. [45]. From $I$ vs. $V - V_g$ measurements, the actual separation between these two charge points is 160 meV, from which we extract a gate capacitance factor (*i.e.*, $V_g = cV$) of $c = 0.08$. The data in Fig. 4a (solid black circles) can be well resolved (continuous red line in Fig. 4a) using the single-level model in Eqn. (7) including two conduction levels (1 and 2) with the following characteristic parameters: $\varepsilon_1 = -25$ meV, $\varepsilon_2 = 135$ meV ($\varepsilon_1 - \varepsilon_2 = 0.160$ eV), $\eta = 0.5$, and $\gamma_{L1} = \gamma_{R1} = 0.21$ meV and $\gamma_{L2} = \gamma_{R2} = 0.06$ meV.

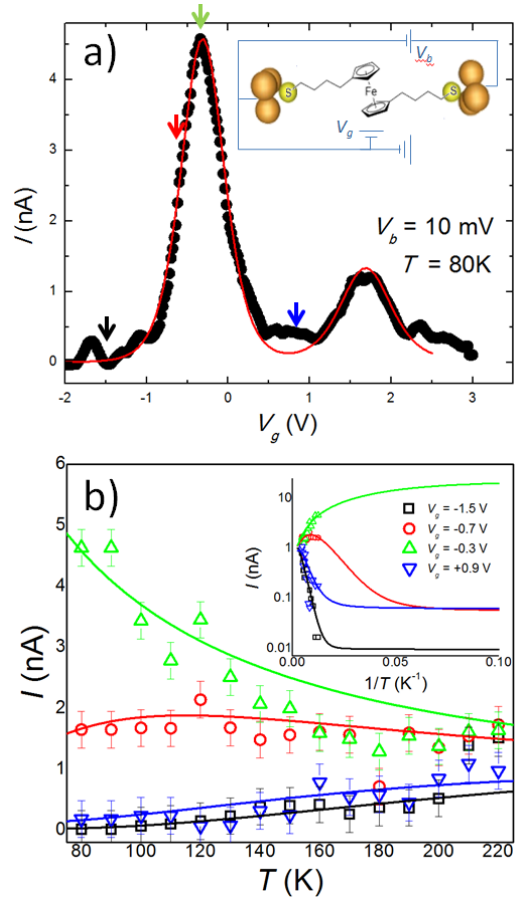

**Figure 4: Temperature dependent tunneling in a single molecule junction**. (**a**) Electrical current through a single S(CH$_2$)$_4$Fc(CH$_2$)$_4$S molecule in a three-terminal SET as a function of gate voltage obtained at bias voltage $V = 10$mV and temperature $T = 80$K. Two charge points at $V_g = -0.3$ and $+1.7$V separate three different charge states in the molecule, *i.e.*, *N*-1, *N* and *N*+1 electrons (Coulomb blockade). *Inset:* Schematic illustration of S(CH$_2$)$_4$Fc(CH$_2$)$_4$S in a three-terminal Al$_2$O$_3$/Au single-electron transistor. (**b**) Experimental (data points) and theoretical (lines) temperature behavior of the tunnel current for four different gate voltages in the temperature range 80-220K (marked with coloured arrows in panel a). The calculated behaviour of the junction was obtained using Eqn. (7) with the following parameters: $\varepsilon_1 = -25$meV, $\varepsilon_2 = 135$ meV ($\varepsilon_1 - \varepsilon_2 = 0.160$ eV), $\eta = 0.5$, and $\gamma_{L1} = \gamma_{R1} = 0.21$ meV and $\gamma_{L2} = \gamma_{R2} = 0.06$ meV.

The measurement in Fig. 4a was repeated at different temperatures up to $T = 220$K. The behavior of the tunnel current with temperature is shown in Fig. 4b for four different gate voltages ($V_g$ = -1.5, -0.7, -0.3, and 0.9V) characteristic of the different conduction regimes in this junction, as indicated by the arrows in Fig. 4a. The agreement with the theoretical expectation is significant for all voltages, whether the conduction is in the Coulomb blockade regime ($V_g$ = -1.5 and +0.9V) or in the transport regime ($V_g$ = -0.7 and − 0.3V).

The inset to Fig. 4b shows fits of current against inverse temperature, from which one can clearly discern the overall behavior for the different conduction regimes, including the transition into a temperature independent regime (quantum coherent plateau) at sufficiently low temperatures. Both the transition temperature and the "activation energy" depend on the gate voltage, as expected from Eqn. (12) and discussed above.

We note that the same temperature behaviour has been observed in several other studies of this family of ferrocene-based molecules in EGaIn junctions [15], illustrating the robustness of the theoretical analysis used in this work to determine the characteristic parameters of solid-state molecular junctions, and serving as a contextualization for the following theoretical analysis of experiments performed in SAM-based molecular junctions.

**A two-level SAM-based molecular junction**

Let us focus now on some very recent experiments in SAM-based molecular junctions where a record-high rectification ratio of three orders of magnitude has been achieved [22]. Figure 5 shows a sketch of the molecular junction formed by a SAM of $S(CH_2)_{11}Fc_2$ molecules (with $Fc_2$ representing a biferrocenylene (Fc=Fc) head group) sandwiched in between GaO$_x^{cond}$/EGaIn (top) and Ag (bottom) electrodes (see Refs. [22,46] for synthesis and experimental details). In these junctions, the Fc$_2$ lies at the end of a long insulating alkyl chain, providing a non-covalent contact (van der Waals coupling) to the top electrode. This results in a highly asymmetric drop of electric potential at both sides of the Fc$_2$ (where the frontier orbitals are located) enabling large rectification ratios. In particular, the Fc$_2$ presents two distinct conduction levels (HOMO and HOMO -1, see Fig. 5) separated by an energy of 0.6 eV, as measured by cyclovoltammetry (CV) in solution and ultra-violet photoemission spectroscopy (UPS) in vacuum, which is much smaller than the HOMO-LUMO gap (which is approx. 2.0-2.5 eV). Figure 5 shows the room-temperature rectification ratio (defined as $R = |I(-V)|/|I(V)|$) as a function of the bias voltage applied to the junction. Apart from the high rectification ratio of $1.1 \times 10^3$ reached at $V = 0.875$ V, the $R$(V) plot shows two clear kinks at around 0.35 and 0.75 volts which were ascribed to the HOMO and HOMO-1 levels subsequently entering the conduction window defined by the difference in electrochemical potential of the top and bottom electrodes [22].

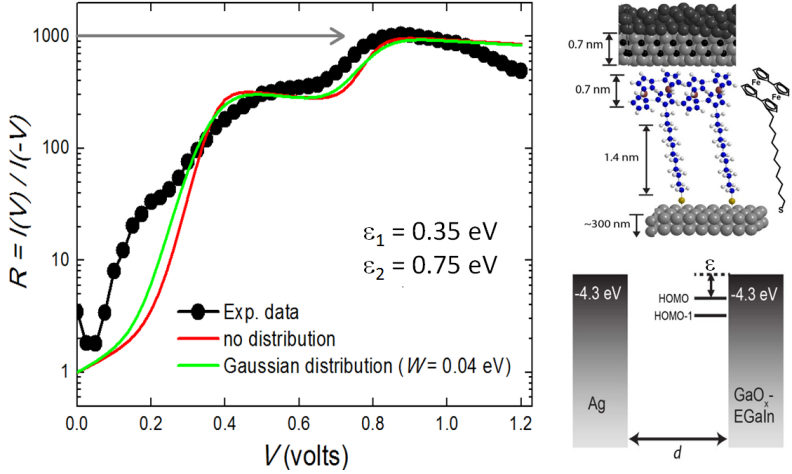

**Figure 5: A two-level molecular junction**. Room-temperature electrical current rectification as a function of bias voltage in a SAM-based junction of $S(CH_2)_{11}Fc_2$ molecules. Two distinct jumps in the rectification at 0.35 and 0.75 volts can be understood in terms of the HOMO and HOMO-1 levels entering the conduction bias window (as depicted in the sketch on the right), after which a record-high rectification ratio of $10^3$ is achieved [22]. The solid lines are fits to a double-level transport model given by Eqns. (5) and (7), with (green) and without (red) the energy distribution of $\varepsilon_1$ and $\varepsilon_2$ given by Eqn. (13).

These results can be easily explained by incorporating a second level into the single-level transport model to account for the two $Fc_2$ levels, *i.e.*, the HOMO and the HOMO-1. Specifically, the addition of two different Lorentzian functions, as given in Eqn. (6), weighted by the corresponding tunneling rates is used in Eqn. (7) to represent the two contributing levels. The results of the calculation using Eqn. (7) are shown in Fig. 5 (continuous red line) for the following characteristic parameters: $\varepsilon_1 = 0.35$ eV, $\varepsilon_2 = 0.75$ eV, $\eta = 0.95$, and $\gamma_{1,L} = 3$ meV, $\gamma_{1,R} = 1$ meV, $\gamma_{2,L} = 5$ meV, $\gamma_{2,R} = 2$ meV. Note: the same results are obtained by Eqn. (5) at room-temperature. The agreement is significant given the approximations in the model and the inherent degree of dispersion in the SAM-based molecular junctions. The theory recovers the voltage position of the two kinks, supporting their association with the HOMO and HOMO-1 levels of the $Fc_2$ group, as well as the overall value of the rectification for most of the bias voltage range, including the maximum rectification of three orders of magnitude. As discussed in Ref. [22], the separation between the levels in the R-V plot (0.4 eV, Fig. 5) is smaller than that measured by CV and UPS (0.6 eV), which can be ascribed to an energy-renormalization of the molecular levels as a result of charge-image effects by the electrodes. This effect should also decrease the HOMO-LUMO gap (estimated to be > 2 eV), allowing it to play some role in the conduction at large bias voltages (reverse polarity), which is likely responsible for the decrease of the rectification ratio above 1 V (see Fig. 5).

Given the fact that molecules are not all identically placed with respect to the electrodes and that their disposition may actually change during the course of a complete measurement, we have used Gaussian distributions for the energies of the HOMO and HOMO-1 levels to account for the degree of dispersion in the system. To do so, we modify Eqn. (7) as follows (note once again that Eqn. (5) can be treated in the same way to give the same results at high temperature):

$$I = \frac{q}{h} \iint_{-\infty}^{\infty} dEdE' \, D_\varepsilon(E) \frac{\gamma_L \gamma_R}{\gamma_L + \gamma_R} [f_L(E) - f_R(E)] g_\varepsilon(E') , \qquad (13)$$

where,

$$g_\varepsilon(E) = Exp\left[\frac{(E-\varepsilon)^2}{2W^2}\right], \qquad (14)$$

The best results are obtained for a width of the Gaussian distribution $W = 0.04$ eV for each level (green line in Fig. 5). This small dispersion in the position of the energy of the molecular levels (<10%) is indicative of the high degree of electrical stability in these SAM-based molecular junctions.

### *Maximum rectification ratio for a single-level junction*

We note that the rectification ratio of three orders of magnitude observed for this molecular junction (with typical level widths $\gamma \sim$ 1-10meV) is at the theoretical maximum limit expected from a single-level transport mechanism for characteristic positions of the frontier orbital with respect to the Fermi energy ($\varepsilon \sim 0.5 - 1 \text{eV}$) and operating voltages ($V \sim 1$ volt). This is easy to deduce by looking in Figure 6 at the current and rectification ratios calculated from Eqn. (7) for a molecular junction, where the parameters have been chosen to be representative of these kinds of junctions (*i.e.*, derived from experimental data). In particular, the calculations have been obtained for a single-level positioned at $\varepsilon = 0.5$ eV, and with a maximally asymmetric voltage divider parameter $\eta = 0$, to maximize the rectification ratio. Obviously, the rectification will be maximum for a bias voltage larger than -0.5 V, which is the forward voltage needed to bring the molecular level into resonance when $\eta = 0$ (as obtained from the definition of $E_\text{m}$ and illustrated in Fig. 2).

Figure 6a shows the behaviour of the room-temperature electrical current with $\gamma_i$ (with $\gamma_L = \gamma_R$) for both polarities of the bias voltage ($V = \pm 1$ volt). As expected, the difference between the forward and reverse currents increases as the tunneling rates (and consequently the level widths) decrease. Both currents reach the same saturation value at sufficiently high tunneling rates (*i.e.*, $\gamma_i \gtrsim 1$ eV), when the level width is larger than the separation between electrochemical potentials in the leads and the conductance becomes constant, as discussed above. The corresponding $R$ for these voltages is shown in Figure 6b for different temperatures, to emphasise that it does not vary below room-temperature. In the range $10^{-5} < \gamma_i \lesssim 1$ eV the electrical rectification decreases exponentially at room-temperature and below, saturating at $R = 1$ (no rectification) for $\gamma_i \gtrsim 1$ eV, corresponding to the behaviour of the current in Fig. 6a. From these results one can conclude that in molecular junctions of this kind, where the $\gamma_i$ takes values typically above 1 meV, the rectification ratio will be limited to three orders of magnitude when a single electrostatic molecular level contributes to the conductance through the junction, which is the value found in the experiments reported in Ref. [22] and marked by the arrows in Fig. 6.

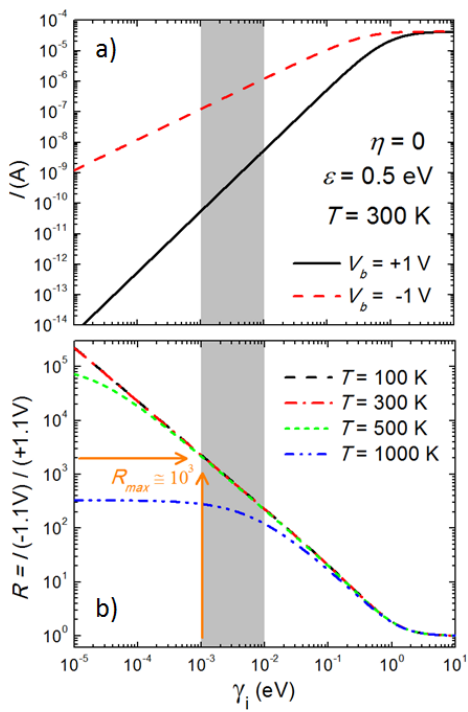

**Figure 6: Limits on Molecular Tunneling Rectification**. (**a**) Electrical current and (**b**) rectification ratio as a function of $\gamma_i$ (with $\gamma_L = \gamma_R$) calculated solving Eqn. (7) for a single-level molecular junction with the following parameters: $\varepsilon = 0.5$ eV and $\eta = 0$. For the typical smallest values of $\gamma_i$ =1 meV, the rectification originated from a single level (*e.g.*, the HOMO orbital) in a molecular junction is theoretically limited to about three orders of magnitude.

We want to stress again that this result is found for voltage conditions that maximize the rectification efficiency (*i.e.*, $\eta = 0$) but with an experimentally relevant energy location of the frontier orbital ($\varepsilon = 0.5$ eV) and typical values of $\gamma_i$. Larger $\varepsilon$ values and, consequently, larger voltages will increase this limit, but not substantially. Evidently, if more than one molecular level is involved in charge transport the rectification can be improved by a factor proportional to the total number of levels involved. However, since it is unlikely to find molecular junctions with more than a few levels in the vicinity of the electrochemical potential of the leads, it is unwise to expect substantially higher electrical rectification ratios originating from the molecular level structure. Imaginative ways should be explored in order to overcome this limit in practice, perhaps by engineering junctions that mix high single-level rectification properties with conformational changes in the molecule that can build up to increase the overall rectification of the junction.

### *Determination of the electrostatic potential profile in a molecular junction*

Some of the authors in this article have recently reported a study of electrical rectification in EGaIn SAM-based junctions where the Fc unit was placed at 14 different positions within the SAM. The molecules have the form of S-$C_n$-Fc-$C_{13-n}$, where $C_n$ represents the number of aliphatic carbons ($CH_2$ or a terminal $CH_3$), with n = 0 to 13 [33]. Controlling the position of the Fc unit (where the molecular HOMO level is localized) along the alkyl chain for different values of n (see sketches in Figure 7) allows us to quantify the rectification response for different energetic symmetries in the junction (values of $\eta$), since the electric potential at both sides of the conduction unit (*i.e.*, Fc) will strongly depend on its position along the chain. This enables sampling of the electrostatic potential profile along the junction gap, which was found to be highly non-linear in this particular molecular junction [33].

Figure 7 shows *R* as a function of n measured at room-temperature and calculated by comparing the electrical currents at $V = \pm 1$V. As can be clearly observed, the rectification behavior is highly non-linear, with several distinct areas, and with the maximum rectification values achieved when the Fc unit is placed close to one of the electrodes (*i.e.*, n = 3 and n > 10), as expected from the asymmetry in the potential drops at both sides of the Fc. For n < 3 the strong hybridization between the Fc and the bottom electrode (chemisorbed contact) leads to large level broadenings and leakage currents, which in turn results in lower rectification ratios (see Ref. [33] for more details). A similar effect (*i.e.*, R saturates), although weaker, arises for n > 10 due to proximity to the top electrode, where the coupling between the molecule and the lead is non-covalent (van der Waals interactions). The intermediate range of Fc positions (4 < $n$ <11) is the most interesting in the context of the analysis of the electrostatic potential profile in this article, since coupling between the Fc unit and the electrodes is non-covalent, and the molecular level responsible for conduction is well defined, with a width that can be considered constant, as we discuss below.

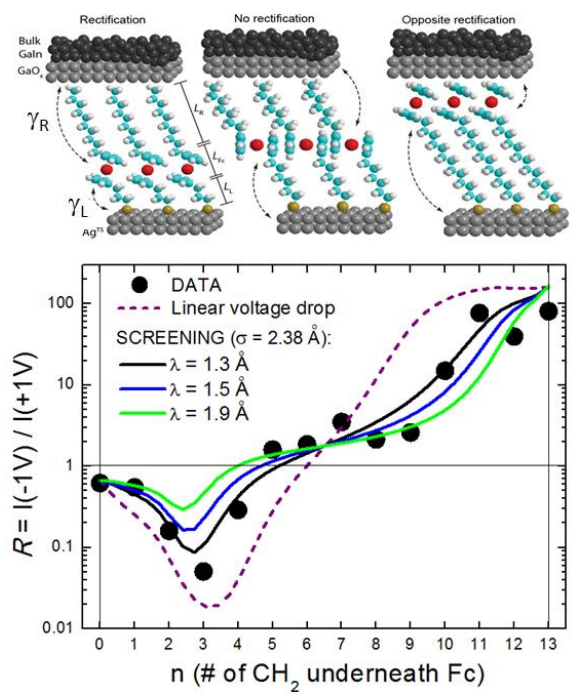

**Figure 7: Electrostatic Potential Profile: Experiment**. **Sketch)** Representation of the EGaIn SAM-based junctions formed by S-$C_n$-Fc-$C_{13-n}$ molecules for three different positions of the Fc unit within the alkyl chain. **Main panel)** Measured rectification ratio of the junctions in Ref. [33] as a function of the position n of the Fc unit within the chain (solid circles). The lines represent fittings to the data using the single-level transport model in Eqn. (7) using different shapes for the electrostatic potential profile in the junction, including linear (dashed purple line) and non-linear as derived from a model that assumes the molecule a cylinder of radius $\sigma$ = 2.38 Å and takes into account screening of the electric field by the molecule (with screening length $\lambda$).

The results in Fig. 7 have been fitted to the single-level model given in Eqn. (7) with characteristic energy parameters defined in the following. We reiterate once again that Eqn. (5) will give the same exact results for this temperature and electric potential conditions. Before getting into the details of the calculations, we emphasise that the fitting results show that the distinct behaviour of *R* for intermediate n values, and particularly its small slope for n = 5 - 9 (Fig. 7), can only be explained by taking into account screening effects, *i.e.*, a non-linear electrostatic potential profile. As we show below, single-level transport models together with a functional treatment to account for electric field screening proposed by Nitzan and collaborators [47] allows determination of the electrostatic potential profile and extraction of a quantitative estimate of the screening length in a molecular junction, which has not been possible to date.

In the following we describe the analytical models that have been employed to fit the rectification data in Figure 7 by means of Eqn. (7). First, for the distance dependence of the energy of the HOMO level, ε, the following "synthetic" function has been employed: $\varepsilon(n) = \varepsilon(n \sim 13) - Ae^{-Bn}$, with $\varepsilon(n \sim 13) = 0.75$ eV (see solid black line in Fig. 8b). The rationale behind this selection is based on the n-dependence of $\varepsilon^{exp}$ obtained from UPS data (see solid data in Fig. 8a), which departs from its linear behavior for distances below n ~ 5. The solid black squares in Fig. 8b, closely following the function used in the calculations (solid black line), represent the result of subtracting a linear function (dashed line in Fig. 8a) from the experimentally obtained value of $\varepsilon^{exp}$. The linear behavior is associated with screening by the metallic electrodes and depends on the Fc-distance (thus the linearity). Therefore, for n > 5 the energy offset ε(n) is expected to be constant (*i.e.*, independent of n). Departures from linearity (as observed for n < 5) are ascribed to a real change in the distance between the molecular level and the Fermi energy of the electrode, which in this case decreases. Figure 8b shows the experimental data of Fig. 8a after subtraction of the linear slope (black squares), together with the function used to fit the data (black line).

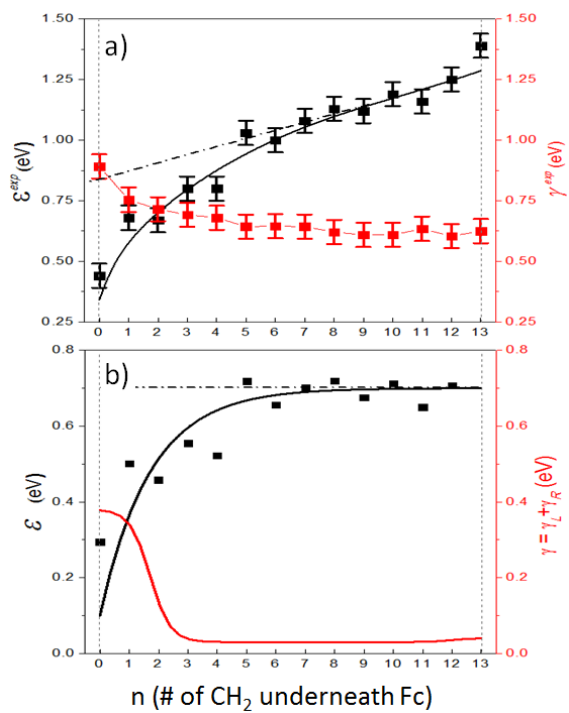

**Figure 8: Molecule-Electrodes coupling**. **a)** Experimental value of $\varepsilon^{exp}$ (left axis) and $\gamma$ of the S-$C_n$-Fc-$C_{13-n}$ SAMs on Ag$^{TS}$ as a function of n obtained from UPS experiments reported in reference [39]. The lines are guides to the eye, with the dashed line representing the linear background expected from screening by the metallic electrodes, which is proportional to the Fc-electrode distance. **b)** Functions employed to represent $\varepsilon$ and $\gamma$ for different positions of the Fc within the alkyl chain (black and red lines, respectively). The solid squares are the data in panel *a* after subtraction of the observed linear slope in the UPS data caused by electron-hole screening effects.

Distance-dependence functions have been chosen to represent the coupling energies $\gamma_L(n)$ and $\gamma_R$ (n), and consequently the HOMO level width $\gamma(n) = \gamma_L(n) + \gamma_R(n)$, and $\varepsilon(n)$ in order to account for the effect of the coupling to the respective electrodes. Figure 8b shows the values of $\gamma(n)$ and $\varepsilon(n)$ used in the calculations (solid curves). The broadening of the HOMO level in proximity with the electrode is expected to increase exponentially with decreasing Fc-electrode distance. Fermi's golden rule gives $\gamma_{L,R} = t_{L,R}^2/D_{L,R}$, where $t_{L,R}$ is the effective coupling matrix and $D_{L,R}$ the electron bandwidth in the electrodes [48]. In general, $\gamma_{L,R} \propto e^{-ekL_{L,R}}$, where $k$ is the tunneling attenuation coefficient, proportional to $\sqrt{E_{barrier}-E_F}$, which for alkyl chains on Au or Ag is on the order of a few eV. As can be observed in Figure 8b, the functions chosen to mimic this effect for $\gamma_L$ and $\gamma_R$ are not purely exponential with n close to the electrodes but saturate for distances below two CH$_2$ units (n < 2 and n > 12) from the respective electrodes. This fitting is based on the spectroscopic data [33] that shows little differences observed in the spectroscopic results between SAMs with n = 0 and 1 (and 2 to some extent). As shown in Figure 8b (red curve), the used dependence for the level width reaches maximum values of $\gamma_L(n<2) = 360$ meV (due to strong hybridization with the Ag electrode) and $\gamma_R(n>11) = 40$ meV, and a base value of $\gamma = 30$ meV (with $\gamma_L = \gamma_R = 15$ meV) for intermediate distances, as obtained from the fitting of the rectification curves (see below). Note that taking the full width at half maximum from UPS data to estimate the molecular level broadening results in an overestimation of $\gamma$ due to limitations of the technique (see Ref. [33] for details) and therefore we only can derive the relative values of $\gamma$.

According to Eqns. (5) and (7), the only other ingredient for calculating the electrical current through the molecular junctions is the functional describing the electrostatic potential profile in the junction, which is represented by $\eta(n)$. We have used two different models to obtain this parameter: i) A simple linear electrostatic potential profile; and, ii) a correction to account for electrical screening in the alkyl chain.

i) *Linear electrostatic potential profile*

Assuming a linear dependence of the voltage drop as a function of the distance between the Fc and the respective electrodes (*i.e.*, $V_L \propto L_R$), one can rewrite $\eta = V_R/(V_L + V_R)$ as $\eta \approx L_R/(L_L + L_R)$.

To account for the voltage drop at the Fc, the expression above can be rewritten assuming a correction $L_{Fc}$ for the respective lengths [49], and the expression for the dimensionless division parameter is:

$$\eta(L_R) = \frac{L_R + L_{Fc}}{L_L + L_R + 2L_{Fc}} + \eta_{shift}, \qquad (14)$$

where $\eta_{shift}$ is an effective shift introduced to account for the asymmetry generated by different couplings of the molecule with the two different electrodes and to explain the small rectification ratios ($R < 2$) observed in equivalent alkyl chains without the Fc unit, or when the Fc is placed in the middle of the chain. The corresponding $\eta$ in the case of our molecule is given in Figure 9 (dashed purple line). Note that 8.7% [= $L_{Fc}/(L_A + 2\,L_{Fc})$] of the voltage drops in the Fc and remains unavailable for use in the rectification process. We note that the strength of this correction is proportional to the size of the active unit employed to sample the electrostatic potential profile, which in our case is substantially larger than the adjacent $CH_2$ units. An ideal unit to sample the electrostatic profile would be one of negligible size (which is not possible, of course). That would also guarantee that the sampled molecule would remain undistorted when moving the active unit within the chain.

The resulting rectification curve obtained using the linear model defined by Eqn. (14) (dashed purple line in Fig. 7) does not give an overall good fit to the data but can explain some features. It quantitatively explains the high rectification ratios for $2 < n < 5$ and $n > 9$ and accounts for the abrupt decrease of rectification for $n < 3$ (resulting from a substantial level broadening due to hybridisation with the left Ag electrode). However, it fails to explain the non-monotonic behaviour of the rectification for intermediate n values, and, particularly, the small slope for $4 < n < 10$ (almost constant). This flat region is one of the main results arising from the sampling procedure to extract the exact shape of the electrostatic potential profile in these junctions, enabled by the capacitive nature of the Fc-electrodes coupling in this range of n values. This discrepancy presents a clear indication that the voltage does not drop linearly within the molecule, as it would be the case in vacuum.

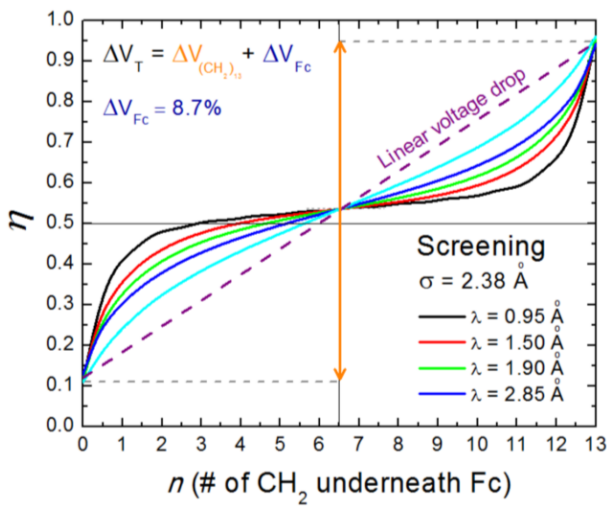

**Figure 9: Electrostatic Potential Profile: Theory**. Electrostatic potential profiles in the molecular junctions as represented by the dimensionless voltage division parameter $\eta$. The curves represent the voltage drop within the alkyl chain on both sides of the Fc unit (totalling 91.3%), which is not the total voltage drop in the junction, since a substantial fraction decays in the Fc (8.7%). Also, the curves have been shifted from $R = 0.5$ in order to account for the different binding energies of the molecule with the left and right electrodes, which leads to a small rectification even in the absence of the Fc unit (or when it is placed at the center of the junction). The linear profile that would be expected in vacuum (red curve) is modified when accounting for electrical screening by the alkyl chain, idealized as a cylinder of diameter $\sigma$ and electrical screening length $\lambda$.

ii) *Correction for the screening of the electric field by the molecule*

The electrochemical potential in electrode-conductor-electrode junctions drops mainly at the contacts, but this is not applicable to the electrostatic potential, which may show profiles extending well into the low-dimensional conductor (*e.g.*, molecule) for sufficiently long electric screening lengths, $\lambda$, which is the case in semiconductors and molecular insulators (see Liang *et al*. [48] and references therein for a more detailed discussion). It is therefore necessary to account for electrical screening effects when trying to explain our observations. For this, we have followed the theoretical formalism developed by Nitzan *et al.* [47], in which the molecule is approximated by a cylindrical conductor of diameter σ, and electric field screening length $\lambda$. As discussed by the authors, the screening length in molecular conductors is an open question, and systems with small HOMO-LUMO gaps are expected to screen well. Assuming that screening over the characteristic length of our system can be described by a Poisson formalism, the voltage division parameter describing the electrostatic potential profile can be written as:

$$\eta(L_R) = \eta_0(L_R) + \frac{A}{\pi}\sum_{m=1}^{\infty}\frac{F_m}{m(1+F_m)}\sin\left(\frac{2\pi m}{L_L+L_R}L_R\right) + \eta_{shift}, \quad (15)$$

where $\eta_0(z)$ is the linear profile given in Eqn. (14). $A$ is a factor to correct for the voltage drop in the Fc. The coefficients $F_m$ can be expressed analyticaly as:

$$F_m = \frac{1}{2}\left(\frac{\sigma}{\lambda}\right)^2 e^{\chi}\int_\chi^\infty du\,\frac{e^{-u}}{u}, \quad (16)$$

with,

$$\chi = \frac{1}{2}\left(2\pi m\frac{\sigma}{\lambda}\right)^2 \quad (17)$$

Using σ = 2.38 Å (= 0.125($L_L + L_R$)), which is taken as the effective diameter of the alkyl chain, the electrostatic potential profiles for different screening lengths ($\lambda$ = 0.95, 1.50, 1.90 and 2.85 Å) are shown in Figure 9. It can be clearly observed that the departure from linear behavior is more pronounced as the screening length decreases. In the limit $\lambda \to 0$, the electrostatic potential will follow the electrochemical potential and drop entirely within the vicinity of the contacts.

The results of the corresponding fittings of the rectification data are shown in Figure 7 for three screening lengths: $\lambda$ = 1.30 Å (black line), $\lambda$ = 1.50 Å (blue line) and $\lambda$ = 1.90 Å (green line). From the good quantitative agreement with the experimental results, where the small slope in the rectification ratio can be well accounted for at intermediated values of n, one can estimate the characteristic screening length to be approximately $\lambda = 1.50 \pm 0.2$ Å. These values of the screening length are comparable to the molecular diameter σ, indicating that the electric field lines from vacuum penetrate well into the molecule.

**DISCUSSION**

We have shown the validity of a simple but formal analytical single-level tunneling model that uses input parameters to calculate *I*(V) curves that can be *experimentally* obtained (tunneling rate γ, the dimensionless division parameter $\eta = V_R/(V_L + V_R)$, and the zero-bias energy offset between the molecular orbital and the electrodes ε). This model was tested against a well-characterized molecular diode and faithfully recovered the rectification ratio and the temperature dependent behavior of the junctions, from which we conclude that it is applicable to describe transport data from both single molecule and SAM-based junctions where intra-molecular collective behavior does not significantly affect the electronic structure of the junction.

One of the main conclusions from our theoretical treatment of the rectification ratio is that the highly non-linear electrostatic potential profile in typical molecular tunneling junctions is the result of a sizeable screening length in the molecule ($\lambda = 1.50 \pm 0.2$ Å, for ferrocene-alkanethiol). Fitting our equations to the experimental results shows that the electrostatics of the molecular junction may be sampled simply by placing the redox Fc unit at different positions along the junction. These results confirm the expectation that potential profiles in insulating molecules are neither linear (as in vacuum) nor do they follow the electrochemical potential profile, which would lead to flat profiles within the entire length of the molecule and would prevent the attainment of high rectification ratios by capacitive coupling with the active transport unit. It is important thus to identify molecular ligands that provide long screening lengths in order to achieve linear-like potential profiles and extend the range of lengths for which the molecular diode would be an efficient rectifier of electrical current.

In addition, we have shown that simple single-level transport models (to describe incoherent and coherent tunneling) can account for the typical temperature dependencies of the conductance in solid state junctions and provide a powerful analysis tool to extract the characteristic parameters governing tunneling in these junctions in a consistent way.

# Supplemental Information

June 2, 2016

**A Single-Level Tunnel Model to Account for Electrical Transport through Single Molecule- and Self-Assembled Monolayer-based Junctions**

by

A. R. Garrigues, L. Yuan, L. Wang, E. R. Mucciolo, D. Thompson, E. del Barco, and C. A. Nijhuis

## 1 The rate equation formulation

Let $M$ be the number of spin-split orbitals in the molecule. The number of possible electronic configurations is equal to $n = 2^M$ (i.e., each orbital can be empty or occupied). Each configuration can be described by a set of occupation numbers $c_\alpha(i)$, with $\alpha = 1, \ldots, n$ and $i = 1, \ldots, M$, where

$$c_\alpha(i) = 0 \text{ or } 1, \tag{1}$$

depending on whether the $i$-th orbital is occupied or not. Call $P_\alpha$ the probability of the $\alpha$ configuration, with $P_\alpha \geq 0$ and $\sum_{\alpha=1}^{n} P_\alpha = 1$.

Under certain approximations, which neglect memory effects and discard correlations between the electrodes and the molecule, the rate equation governing the change in the configuration probabilities over time can be expressed as [1,2,3]

$$\frac{dP_\alpha}{dt} = -P_\alpha \sum_{\beta \neq \alpha} \Gamma_{\alpha \to \beta} + \sum_{\beta \neq \alpha} \Gamma_{\beta \to \alpha} P_\beta, \tag{2}$$

where $\Gamma_{\alpha \to \beta}$ is the rate of the $\alpha \to \beta$ transition. Equation (2) is the well-known Pauli master equation [4]. We note here that there are several ways to derive this equation from fundamental theories, such as nonequilibrium quantum statistical mechanics [5,6,7], as well as from a semiclassical Boltzmann kinetic equation [1]. Here, rather than repeating those standard derivations, we focus on the use of the rate equation to compute transport proeperties in the context of the experiments reported in the main text. The main goal is to obtain expressions for the charge current across the molecule in different asymptotic regimes.

We can express Eq. (2) in matrix form,

$$\frac{dP_\alpha}{dt} = \sum_\beta \Lambda_{\alpha\beta} P_\beta, \tag{3}$$

where

$$\Lambda_{\alpha\beta} = \begin{cases} -\sum_{\gamma \neq \alpha} \Gamma_{\alpha \to \gamma}, \text{if } \alpha = \beta \\ \Gamma_{\beta \to \alpha}, \text{if } \alpha \neq \beta \end{cases}. \tag{4}$$

Notice that, consistent with the normalization condition, we have

$$\sum_{\alpha=1}^n \frac{dP_\alpha}{dt} = -\sum_{\alpha=1}^n P_\alpha \sum_{\beta \neq \alpha} \Gamma_{\alpha \to \beta} + \sum_{\alpha=1}^n \sum_{\beta \neq \alpha} \Gamma_{\beta \to \alpha} P_\beta \tag{5}$$

$$= -\sum_{\alpha=1}^n P_\alpha \sum_{\beta \neq \alpha} \Gamma_{\alpha \to \beta} + \sum_{\beta=1}^n P_\beta \sum_{\alpha \neq \beta} \Gamma_{\beta \to \alpha} \tag{6}$$

$$= 0, \tag{7}$$

implying that

$$\sum_{\alpha=1}^n \Lambda_{\alpha\beta} = 0, \tag{8}$$

as it can be easily verified. On the other hand,

$$\sum_{\beta=1}^n \Lambda_{\alpha\beta} = -\sum_{\gamma \neq \alpha} \Gamma_{\alpha \to \gamma} + \sum_{\beta \neq \alpha} \Gamma_{\beta \to \alpha} \tag{9}$$

$$= \sum_{\beta \neq \alpha} (\Gamma_{\beta \to \alpha} - \Gamma_{\alpha \to \beta}). \tag{10}$$

Thus, if $\Gamma_{\beta \to \alpha} = \Gamma_{\alpha \to \beta}$ for all transitions, then the r.h.s. of this equation is equal to zero. This means that a trivial stationary solution exits where $P_\alpha = 1/n$ for all $\alpha = 1, \ldots, n$. However, when $\Gamma_{\beta \to \alpha} \neq \Gamma_{\alpha \to \beta}$, other nontrivial stationary solution exist as well.

## 2 Stationary solutions

In order to obtain a stationary solution to the rate equations set

$$\frac{dP_\alpha}{dt} = 0 \tag{11}$$

for all $\alpha = 1, \ldots, n$. This implies

$$\sum_\beta \Lambda_{\alpha\beta} P_\beta = 0. \tag{12}$$

Thus, to find the set of stationary probabilities $\{P_\alpha\}$, one needs to find the right eigenvector corresponding to the zero eigenvalue of the matrix $\Lambda$.

## 3 Transition rates

Let $E_\alpha$ be the total energy of the molecule and $N_\alpha$ be the total number of electrons in the $\alpha$ configuration. When an electron hops from one of the leads and into the molecule, energy conservation requires

$$\varepsilon + E_\alpha = E_\beta, \tag{13}$$

where $\alpha(\beta)$ is the molecule's configuration before(after) the hopping and $\varepsilon$ is the energy of the electronic state in the lead. For this transition to take place, the state with energy $\varepsilon$ in the lead must have a finite occupation number, namely $f\left(\frac{\varepsilon-\mu_l}{k_B T}\right) > 0$, where $\mu_l$ is the lead's chemical potential ($l = R, L$) and $T$ is the temperature. Here, $f(x)$ denotes the Fermi-Dirac distribution,

$$f(x) = \frac{1}{e^x + 1}. \tag{14}$$

In the opposite case, when an electron hops from the molecule and into the lead, we have

$$E_\alpha = \varepsilon + E_\beta. \tag{15}$$

Now, the occupation number of the state with energy $\varepsilon$ must be such that $f\left(\frac{\varepsilon-\mu_l}{k_B T}\right) < 1$.

Call $\gamma_R$ and $\gamma_L$ the level widths due to the coupling to the right and left leads, respectively. We can split the transition rate into two contributions,

$$\Gamma_{\alpha\to\beta} = \Gamma^R_{\alpha\to\beta} + \Gamma^L_{\alpha\to\beta}, \tag{16}$$

where [1,2,3]

$$\Gamma^l_{\alpha\to\beta} = \begin{cases} (\gamma_l/\hbar) f\left(\frac{E_\beta - E_\alpha - \mu_l}{k_B T}\right), & \text{if } N_\beta = N_\alpha + 1 \text{ and } d_H(c_\beta, c_\alpha) = 1 \\ (\gamma_l/\hbar)\left[1 - f\left(\frac{E_\alpha - E_\beta - \mu_l}{k_B T}\right)\right], & \text{if } N_\beta = N_\alpha - 1 \text{ and } d_H(c_\beta, c_\alpha) = 1 \\ 0, \text{otherwise} \end{cases}, l = R, L, \tag{17}$$

where $d_H(c_\beta, c_\alpha)$ is the Hamming distance between the binary sets $c_\beta$ and $c_\alpha$. Namely, only transitions where the number of electrons in the molecule changes by one are allowed. Notice that the bias voltage is equal to $V = (\mu_L - \mu_R)/e$, where $e$ denotes the electron charge. Equation (17) can be derived in a number of ways, with the most standard being Fermi's Golden Rule or time-dependent perturbation theory on the level widths $\gamma_R$ and $\gamma_L$.

The rate equation approach is valid when $k_B T \gg \gamma_{R,L}$ (since it assumes a perfectly sharp energy lever in the molecule), and when the tunneling through the molecule is sequentially incoherent. Below, we discuss the validity of the rate equation in more detail.

Implicit in Eq. (17) is the assumption that the molecule's energy levels are sharp, such that $\gamma_R, \gamma_L$ are much smaller than other energy scales of the problem, such as $k_B T$, $eV$, and the separation of energy levels in the molecule.

## 4 Current

The current coming from the left lead is equal to

$$I_L = -e \sum_{\alpha,\beta} \Delta N_{\alpha\to\beta} \Gamma^L_{\alpha\to\beta} P_\alpha, \tag{18}$$

where $\Delta N_{\alpha\to\beta} = N_\beta - N_\alpha$.

## 5 Total energy

The total energy in the molecule can be broken down as follows (constant charging energy model):

$$E_\alpha = \frac{1}{2} N_\alpha (N_\alpha - 1) E_c - e V_g N_\alpha + \sum_{i=1}^{M} c_\alpha(i) \varepsilon_i, \qquad (19)$$

where $E_c$ is the charging energy, $V_g$ is the gate voltage, and $\{\varepsilon_i\}_{i=1,\ldots,M}$ are the energies of the orbitals.

## 6 Single-level case

Let us apply this formulation compute the stationary current of a molecule with a single orbital (spinless case), in which case $M = 1$ and $n = 2$. $P_{0(1)}$ corresponds to the probability of the empty(filled) state. The stationary problem is defined by the matrix

$$\Lambda = \begin{pmatrix} -\Gamma_{0\to 1} & \Gamma_{1\to 0} \\ \Gamma_{0\to 1} & -\Gamma_{1\to 0} \end{pmatrix}, \qquad (20)$$

where

$$\Gamma_{0\to 1} = \Gamma^R_{0\to 1} + \Gamma^L_{0\to 1} \qquad (21)$$

and

$$\Gamma_{1\to 0} = \Gamma^R_{1\to 0} + \Gamma^L_{1\to 0}, \qquad (22)$$

with

$$\Gamma^l_{0\to 1} = (\gamma_l/\hbar) f\left(\frac{E_1 - E_0 - \mu_l}{k_B T}\right), l = L, R, \qquad (23)$$

and

$$\Gamma^l_{1\to 0} = (\gamma_l/\hbar)\left[1 - f\left(\frac{E_1 - E_0 - \mu_l}{k_B T}\right)\right], l = L, R. \qquad (24)$$

The eigenvector of the $\Lambda$ matrix with zero eigenvalue corresponds to

$$P_0 = \frac{\Gamma_{1\to 0}}{\Gamma_{1\to 0} + \Gamma_{0\to 1}} = \frac{\hbar \Gamma_{1\to 0}}{\gamma_R + \gamma_L} \qquad (25)$$

and

$$P_1 = \frac{\Gamma_{0\to 1}}{\Gamma_{1\to 0} + \Gamma_{0\to 1}} = \frac{\hbar \Gamma_{0\to 1}}{\gamma_R + \gamma_L}. \qquad (26)$$

The current coming from the left lead is equal to

$$I_L = -e(P_0 \Gamma^L_{0 \to 1} - P_1 \Gamma^L_{1 \to 0}) \tag{27}$$

$$= -\frac{e\hbar}{\gamma_R + \gamma_L}(\Gamma_{1 \to 0}\Gamma^L_{0 \to 1} - \Gamma_{0 \to 1}\Gamma^L_{1 \to 0}) \tag{28}$$

$$= -\frac{e\hbar}{\gamma_R + \gamma_L}(\Gamma^R_{1 \to 0}\Gamma^L_{0 \to 1} - \Gamma^R_{0 \to 1}\Gamma^L_{1 \to 0}) \tag{29}$$

$$= -\frac{e}{\hbar}\frac{\gamma_R \gamma_L}{\gamma_R + \gamma_L}\left\{\left[1 - f\left(\frac{E_{10} - \mu_R}{k_B T}\right)\right] f\left(\frac{E_{10} - \mu_L}{k_B T}\right) - f\left(\frac{E_{10} - \mu_R}{k_B T}\right) \right.$$
$$\left. \times \left[1 - f\left(\frac{E_{10} - \mu_L}{k_B T}\right)\right]\right\} \tag{30}$$

$$= -\frac{e}{\hbar}\frac{\gamma_R \gamma_L}{\gamma_R + \gamma_L}\left[f\left(\frac{E_{10} - \mu_L}{k_B T}\right) - f\left(\frac{E_{10} - \mu_R}{k_B T}\right)\right], \tag{31}$$

where $E_{10} = E_1 - E_0 = -eV_g + \varepsilon_1$.

## 6.1 Finite level width

When the broadening of the energy level is not negligible, we have to modify the calculations to account for the uncertainty in $\epsilon_1$. Let $\gamma$ be the total level width and $D_1(\varepsilon)$ the density of state profile associated to the single-level configuration; for instance, consider the Lorentzian profile

$$D_1(\varepsilon) = \frac{1}{\pi}\frac{\gamma/2}{(\varepsilon - \varepsilon_1)^2 + (\gamma/2)^2}, \tag{32}$$

with $\int d\varepsilon D_1(\varepsilon) = 1$. The modified expressions for the transition rates are

$$\Gamma^l_{0 \to 1} = (\gamma_l/\hbar) \int d\varepsilon D_1(\varepsilon) f\left(\frac{E_1 - E_0 - \mu_l}{k_B T}\right), l = L, R, \tag{33}$$

and

$$\Gamma^l_{1 \to 0} = (\gamma_l/\hbar) \int d\varepsilon D_1(\epsilon) \left[1 - f\left(\frac{E_1 - E_0 - \mu_l}{k_B T}\right)\right], l = L, R, \tag{34}$$

where

$$E_1 - E_0 = -eV_g + \varepsilon. \tag{35}$$

Notice that

$$\Gamma^R_{0 \to 1} + \Gamma^R_{1 \to 0} = (\gamma_R/\hbar) \int d\varepsilon D_1(\varepsilon) = \gamma_R/\hbar \tag{36}$$

and

$$\Gamma^L_{0 \to 1} + \Gamma^L_{1 \to 0} = (\gamma_L/\hbar) \int d\varepsilon D_1(\varepsilon) = \gamma_L/\hbar. \tag{37}$$

Therefore,

$$\Gamma_{0 \to 1} + \Gamma_{1 \to 0} = (\gamma_R + \gamma_L)/\hbar, \tag{38}$$

$$P_0 = \frac{1}{\gamma_R + \gamma_L} \int d\varepsilon D_1(\varepsilon) \sum_l \gamma_l \left[1 - f\left(\frac{E_1 - E_0 - \mu_l}{k_B T}\right)\right] \quad (39)$$

$$= \frac{1}{\gamma_R + \gamma_L} \sum_l \gamma_l \int d\epsilon \varepsilon D_1(\varepsilon + eV_g) \left[1 - f\left(\frac{\varepsilon - \mu_l}{k_B T}\right)\right] \quad (40)$$

and

$$P_1 = \frac{1}{\gamma_R + \gamma_L} \int d\varepsilon D_1(\varepsilon) \sum_l \gamma_l f\left(\frac{E_1 - E_0 - \mu_l}{k_B T}\right) \quad (41)$$

$$= \frac{1}{\gamma_R + \gamma_L} \sum_l \gamma_l \int d\varepsilon D_1(\varepsilon + eV_g) f\left(\frac{\varepsilon - \mu_l}{k_B T}\right). \quad (42)$$

Going back to the expression defining the current through the left lead, we find

$$I_L = -e(P_0 \Gamma^L_{0\to 1} - P_1 \Gamma^L_{1\to 0}) \quad (43)$$

$$= -\frac{e}{\hbar} \frac{1}{\gamma_R + \gamma_L} (\Gamma_{1\to 0} \Gamma^L_{0\to 1} - \Gamma_{0\to 1} \Gamma^L_{1\to 0}) \quad (44)$$

$$= -\frac{e}{\hbar} \frac{1}{\gamma_R + \gamma_L} [(\Gamma^R_{1\to 0} + \Gamma^L_{1\to 0})\Gamma^L_{0\to 1} - (\Gamma^R_{0\to 1} + \Gamma^L_{0\to 1})\Gamma^L_{1\to 0}] \quad (45)$$

$$= -\frac{e}{\hbar} \frac{1}{\gamma_R + \gamma_L} [\Gamma^R_{1\to 0} \Gamma^L_{0\to 1} - \Gamma^R_{0\to 1} \Gamma^L_{1\to 0}] \quad (46)$$

$$= -\frac{e}{\hbar} \frac{\gamma_R \gamma_L}{\gamma_R + \gamma_L} \int d\varepsilon D_1(\varepsilon) \int d\varepsilon' D_1(\varepsilon') \left\{ \left[1 - f\left(\frac{-eV_g + \varepsilon - \mu_R}{k_B T}\right)\right] f\left(\frac{-eV_g + \varepsilon' - \mu_L}{k_B T}\right) \right.$$
$$\left. - f\left(\frac{-eV_g + \varepsilon - \mu_R}{k_B T}\right) \left[1 - f\left(\frac{-eV_g + \varepsilon' - \mu_L}{k_B T}\right)\right] \right\} \quad (47)$$

$$= -\frac{e}{\hbar} \frac{\gamma_R \gamma_L}{\gamma_R + \gamma_L} \int d\varepsilon D_1(\varepsilon) \int d\varepsilon' D_1(\varepsilon') \left\{ f\left(\frac{-eV_g + \varepsilon' - \mu_L}{k_B T}\right) - f\left(\frac{-eV_g + \varepsilon - \mu_R}{k_B T}\right) \right\} \quad (48)$$

$$= -\frac{e}{\hbar} \frac{\gamma_R \gamma_L}{\gamma_R + \gamma_L} \int d\varepsilon D_1(\varepsilon) \left[ f\left(\frac{-eV_g + \varepsilon - \mu_L}{k_B T}\right) - f\left(\frac{-eV_g + \varepsilon - \mu_R}{k_B T}\right) \right] \quad (49)$$

$$= -\frac{e}{\hbar} \frac{\gamma_R \gamma_L}{\gamma_R + \gamma_L} \int d\varepsilon D_1(\varepsilon + eV_g) \left[ f\left(\frac{\varepsilon - \mu_L}{k_B T}\right) - f\left(\frac{\varepsilon - \mu_R}{k_B T}\right) \right] \quad (50)$$

$$= -\frac{e}{h} \frac{\gamma_R \gamma_L}{\gamma_R + \gamma_L} \int d\varepsilon \frac{\gamma}{(\varepsilon - \varepsilon_1 + eV_g)^2 + (\gamma/2)^2} \left[ f\left(\frac{\varepsilon - \mu_L}{k_B T}\right) - f\left(\frac{\varepsilon - \mu_R}{k_B T}\right) \right]. \quad (51)$$

This expression generalizes the previous one, Eq. (31), to include a finite level width $\gamma$. It is straightforward to check that Eq. (51) recoves Eq. (31) when $\gamma \to 0$.

It is natural to assume that the total level width can be broken into three components,

$$\gamma = \gamma_R + \gamma_L + \gamma_0, \quad (52)$$

where $\gamma_0$ represents the broadening caused by effects other than the leakage of charge through the leads.

## 6.2 Adding spin

To add spin, we split the configuration where the molecule level is occupied into two ($\uparrow, \downarrow$), resulting in a total of three configurations: $i = 0, \uparrow, \downarrow$ (we forbid double occupancy by assuming that the charging energy $E_c$ is a very large energy scale, namely, $E_c \gg k_B T, eV, |\varepsilon_1|$). Let us assume that the molecular level is spin degenerate. Then, the total current through the left lead is given by the expression

$$I_L = -e[P_0(\Gamma^L_{0\to\uparrow} + \Gamma^L_{0\to\downarrow}) - P_\uparrow \Gamma^L_{\uparrow\to 0} - P_\downarrow \Gamma^L_{\downarrow\to 0}]. \tag{53}$$

The rate equations are

$$\frac{dP_0}{dt} = -P_0(\Gamma_{0\to\uparrow} + \Gamma_{0\to\downarrow}) + P_\uparrow \Gamma_{\uparrow\to 0} + P_\downarrow \Gamma_{\downarrow\to 0}, \tag{54}$$

$$\frac{dP_\uparrow}{dt} = -P_\uparrow \Gamma_{\uparrow\to 0} + P_0 \Gamma_{0\to\uparrow}, \tag{55}$$

$$\frac{dP_\downarrow}{dt} = -P_\downarrow \Gamma_{\downarrow\to 0} + P_0 \Gamma_{0\to\downarrow}. \tag{56}$$

Solving for the steady state yields

$$P_0 = \frac{\Gamma_{\uparrow\to 0}\Gamma_{\downarrow\to 0}}{\Gamma_{\uparrow\to 0}\Gamma_{\downarrow\to 0} + \Gamma_{0\to\uparrow}\Gamma_{\downarrow\to 0} + \Gamma_{0\to\downarrow}\Gamma_{\uparrow\to 0}}, \tag{57}$$

$$P_\uparrow = \frac{\Gamma_{0\to\uparrow}\Gamma_{\downarrow\to 0}}{\Gamma_{\uparrow\to 0}\Gamma_{\downarrow\to 0} + \Gamma_{0\to\uparrow}\Gamma_{\downarrow\to 0} + \Gamma_{0\to\downarrow}\Gamma_{\uparrow\to 0}}, \tag{58}$$

and

$$P_\downarrow = \frac{\Gamma_{0\to\downarrow}\Gamma_{\uparrow\to 0}}{\Gamma_{\uparrow\to 0}\Gamma_{\downarrow\to 0} + \Gamma_{0\to\uparrow}\Gamma_{\downarrow\to 0} + \Gamma_{0\to\downarrow}\Gamma_{\uparrow\to 0}}. \tag{59}$$

Assuming spin degeneracy in the leads, we find

$$\Gamma^l_{0\to\uparrow} = \Gamma^l_{0\to\downarrow} = (\gamma_l/\hbar)\int d\varepsilon D_1(\varepsilon) f\left(\frac{E_1 - E_0 - \mu_l}{k_B T}\right) \equiv \Gamma^l_{0\to 1}, l = L, R, \tag{60}$$

and

$$\Gamma^l_{\uparrow\to 0} = \Gamma^l_{\downarrow\to 0} = (\gamma_l/\hbar)\int d\varepsilon D_1(\varepsilon)\left[1 - f\left(\frac{E_1 - E_0 - \mu_l}{k_B T}\right)\right] \equiv \Gamma^l_{1\to 0}, l = L, R. \tag{61}$$

Therefore,

$$\Gamma_{0\to\uparrow} = \Gamma_{0\to\downarrow} = \int d\varepsilon D_1(\varepsilon)[\gamma_R f_R(\varepsilon) + \gamma_L f_L(\varepsilon)] \equiv \Gamma_{0\to 1} \tag{62}$$

and

$$\Gamma_{\uparrow\to 0} = \Gamma_{\downarrow\to 0} = \int d\varepsilon D_1(\varepsilon)[\gamma_R + \gamma_L - \gamma_R f_R(\varepsilon) - \gamma_L f_L(\varepsilon)]/\hbar \tag{63}$$

$$= (\gamma_R + \gamma_L)/\hbar - \int d\varepsilon D_1(\varepsilon)[\gamma_R f_R(\varepsilon) + \gamma_L f_L(\varepsilon)]/\hbar \tag{64}$$

$$= (\gamma_R + \gamma_L)/\hbar - \Gamma_{0\to 1} \equiv \Gamma_{1\to 0}, \tag{65}$$

where we introduced

$$f_l(\varepsilon) = f\left(\frac{E_1 - E_0 - \mu_l}{k_B T}\right). \tag{66}$$

Notice that rates and probabilities do not depend on spin. Thus, we can recast the problem in terms of $P_0$ and $P_1 = P_\uparrow + P_\downarrow$. Then, we find

$$P_0 = \frac{\Gamma_{1\to 0}}{\Gamma_{1\to 0} + 2\Gamma_{0\to 1}} \tag{67}$$

and

$$P_1 = \frac{2\Gamma_{0\to 1}}{\Gamma_{1\to 0} + 2\Gamma_{0\to 1}}. \tag{68}$$

Plugging them into the expression for the current, we get

$$I_L = -e(2P_0 \Gamma^L_{0\to 1} - P_1 \Gamma^L_{1\to 0}) \tag{69}$$

$$= -2e\frac{\Gamma_{1\to 0}\Gamma^L_{0\to 1} - \Gamma_{0\to 1}\Gamma^L_{1\to 0}}{\Gamma_{1\to 0} + 2\Gamma_{0\to 1}} \tag{70}$$

$$= -2e\frac{\Gamma^R_{1\to 0}\Gamma^L_{0\to 1} - \Gamma^R_{0\to 1}\Gamma^L_{1\to 0}}{(\gamma_R + \gamma_L)/\hbar + \Gamma_{0\to 1}} \tag{71}$$

$$= -\frac{2e}{h}\frac{\gamma_R \gamma_L}{(\gamma_R + \gamma_L) + \hbar\Gamma_{0\to 1}} \int d\varepsilon \frac{\gamma}{(\varepsilon - \varepsilon_1 + eV_g)^2 + (\gamma/2)^2}\left[f\left(\frac{\varepsilon - \mu_L}{k_B T}\right) - f\left(\frac{\varepsilon - \mu_R}{k_B T}\right)\right] \tag{72}$$

where

$$\Gamma_{0\to 1} = \int d\varepsilon \frac{\gamma}{(\varepsilon - \varepsilon_1 + eV_g)^2 + (\gamma/2)^2}\left[\gamma_R f\left(\frac{\varepsilon - \mu_L}{k_B T}\right) + \gamma_L f\left(\frac{\varepsilon - \mu_R}{k_B T}\right)\right]/\hbar. \tag{73}$$

Notice that because of the term $\Gamma_{0\to 1}$ in the denominator of the prefactor in Eq. (72), the expression for the current in the presence of spin is not exactly equal to twice that for the spinless case. However, if we are only interested in linear response, we can set $\mu_L = \mu_R = \mu$ in Eq. (73), in which case we obtain $\Gamma_{0\to 1} \approx (\gamma_R + \gamma_L)/\hbar$, provided that $\varepsilon_1 - eV_g < \mu$ (namely, when the energy level is brought below the Fermi energy in the leads). Then, the factor of 2 is approximately cancelled and we recover the expression for the spinless current. The current for the spinfull case is only exactly equal to twice that for the spinless case when the charging energy in the molecule is zero (non-interacting limit), in which case conductance through the molecule is spin degenerate.

# 7  Exact solution of the single-level case (spinless)

It is possible to solve exactly the fully coherent single-level case by using the Keldysh non-equilibrium technique [7], or even scattering theory, since no many-body interactions are present [8]. The result is the following: the probability of the level to be occupied is equal to

$$P_1 = \sum_l \frac{\gamma_l}{\gamma_R + \gamma_L} \int \frac{d\varepsilon}{2\pi} f\left(\frac{\varepsilon - \mu_l}{k_B T}\right) \frac{\gamma}{(\varepsilon - \varepsilon_1 + eV_g)^2 + (\gamma/2)^2}, \tag{74}$$

where $\gamma = \gamma_R + \gamma_L$ (absence of any level broadening other than leakage of charge through the leads). The probability of the empty level configuration is $P_0 = 1 - P_1$. The expressions for the probabilities are identical to those obtained with the rate equations after the broadening of the energy level is incorporated.

The fact that the coherent and incoherent formulations yield the same results for the probabilities is not surprising. For single channel leads and a single level in the molecule, interference plays no role since there is only one conduction path. When the molecule has multiple independent paths for electrons to hop in and out, then the coherent and incoherent predictions depart, since interference between paths can result in enhancement or depletion of certain configuration occupations.

An expression for the current was derived by Jauho, Wingreen, and Meir [9] using the Keldysh Green's function technique. Their result is

$$I_L = -\frac{e}{h}\int d\varepsilon \frac{\gamma_R \gamma_L}{(\varepsilon - \varepsilon_1 + eV_g)^2 + (\gamma/2)^2}\left[f\left(\frac{\varepsilon - \mu_L}{k_B T}\right) - f\left(\frac{\varepsilon - \mu_R}{k_B T}\right)\right]. \tag{75}$$

Contrary to our previous derivation using rate equations, this expression fully takes into account coherence. Yet, Eq. (75) and Eq. (51) are identical, provided that we set $\gamma = \gamma_R + \gamma_L$ (namely, no level broadening other than that due leakage through the leads). To some extend this should come as a surprise, as the coherent transport formulation contains incoherent, sequential regime as a limit.

Notice that for the spinfull case, one simply need to insert a factor of 2 on the right-hand-side of Eq. (75).

**Asymptotic limits for the current**

Notice that

$$f\left(\frac{\varepsilon - \mu_L}{k_B T}\right) - f\left(\frac{\varepsilon - \mu_R}{k_B T}\right) = \frac{\sinh\left(\frac{eV_b}{2k_B T}\right)}{\cosh\left(\frac{\varepsilon + E_F}{k_B T}\right) + \cosh\left(\frac{eV_b}{2k_B T}\right)}, \tag{76}$$

where $eV_b = \mu_L - \mu_R$ and $E_F = (\mu_L + \mu_R)/2$. Defining $\varepsilon' = \varepsilon - \varepsilon_1 + eV_g$, we can then rewrite Eq. (75) as

$$I_L = -\frac{e}{\hbar}\frac{\gamma_R \gamma_L}{\gamma_R + \gamma_L}\int d\varepsilon' D(\varepsilon')\left[\frac{\sinh\left(\frac{eV_b}{2k_B T}\right)}{\cosh\left(\frac{\varepsilon' + \varepsilon_1 - eV_g + E_F}{k_B T}\right) + \cosh\left(\frac{eV_b}{2k_B T}\right)}\right], \tag{77}$$

where

$$D(\varepsilon') = \frac{1}{\pi}\frac{\gamma/2}{\varepsilon'^2 + (\gamma/2)^2}. \tag{78}$$

It is easy to show that

$$\int d\varepsilon' D(\varepsilon') = 1. \tag{79}$$

Without loss of generality, we can set $eV_g = E_F$. Thus, $\varepsilon_1$ becomes the position of the energy level with respect the Fermi energy in the leads at zero bias. Then,

$$I_L = -\frac{e}{\hbar}\frac{\gamma_R\gamma_L}{\gamma_R+\gamma_L}\sinh\left(\frac{eV_b}{2k_BT}\right)\int d\varepsilon' D(\varepsilon')\frac{1}{\cosh\left(\frac{\varepsilon'+\varepsilon_1}{k_BT}\right)+\cosh\left(\frac{eV_b}{2k_BT}\right)}. \tag{80}$$

Let us look at some asymptotic limits.

- $\gamma \ll e|V_b| \ll k_BT \ll |\varepsilon_1|$: Weak broadening, finite bias, large temperature.

$$\begin{align}
I_L &\approx -\frac{e}{\hbar}\frac{\gamma_R\gamma_L}{\gamma_R+\gamma_L}\left(\frac{eV_b}{2k_BT}\right)\int d\varepsilon' D(\varepsilon')\frac{1}{\cosh\left(\frac{\varepsilon'+\varepsilon_1}{k_BT}\right)+1} \tag{81}\\
&\approx -\frac{e}{\hbar}\frac{\gamma_R\gamma_L}{\gamma_R+\gamma_L}\left(\frac{eV_b}{k_BT}\right)e^{-\varepsilon_1/k_BT}. \tag{82}
\end{align}$$

The current in this case shows an activation behavior, with the activation energy being the offset between the energy level in the molecule and the Fermi energy in the leads. Linear bias regime. On resonance, we find

$$I_L \approx -\frac{e}{\hbar}\frac{\gamma_R\gamma_L}{\gamma_R+\gamma_L}\left(\frac{eV_b}{k_BT}\right). \tag{83}$$

- $\gamma \ll |\epsilon_1| \ll k_BT \ll e|V_b|$: Weak broadening, intermediate temperature, large bias, nearly on resonance.

$$\begin{align}
I_L &\approx -\frac{e}{\hbar}\frac{\gamma_R\gamma_L}{\gamma_R+\gamma_L}\sinh\left(\frac{eV_b}{2k_BT}\right)\int d\varepsilon' D(\varepsilon')\frac{1}{1+\cosh\left(\frac{eV_b}{2k_BT}\right)} \tag{84}\\
&\approx -\frac{e}{\hbar}\frac{\gamma_R\gamma_L}{\gamma_R+\gamma_L}. \tag{85}
\end{align}$$

The current is approximatelly temperature and bias independent (non-linear bias regime).

- $\gamma \ll k_BT \ll |\epsilon_1| \ll e|V_b|$: Similar to the previous case, more off-resonance.

$$\begin{align}
I_L &\approx -\frac{e}{\hbar}\frac{\gamma_R\gamma_L}{\gamma_R+\gamma_L}\frac{e^{e|V_b|/2k_BT}}{2}\int d\varepsilon' D(\varepsilon')\frac{1}{\cosh\left(\frac{\varepsilon'+\varepsilon_1}{k_BT}\right)+e^{e|V_b|/2k_BT}/2} \tag{86}\\
&\approx -\frac{e}{\hbar}\frac{\gamma_R\gamma_L}{\gamma_R+\gamma_L}. \tag{87}
\end{align}$$

The current is again approximatelly temperature and bias independent.

- $\gamma \ll k_BT \ll e|V_b| < |\epsilon_1|$: Similar to the previous case, but even more off-resonance.

$$\begin{align}
I_L &\approx -\frac{e}{\hbar}\frac{\gamma_R\gamma_L}{\gamma_R+\gamma_L}\frac{e^{e|V_b|/2k_BT}}{2}\int d\varepsilon' D(\varepsilon')\frac{1}{\cosh\left(\frac{\varepsilon'+\varepsilon_1}{k_BT}\right)+e^{e|V_b|/2k_BT}/2} \tag{88}\\
&\approx -\frac{e}{\hbar}\frac{\gamma_R\gamma_L}{\gamma_R+\gamma_L}e^{(e|V_b|/2-|\varepsilon_1|)/k_BT}. \tag{89)
\end{align}$$

The current shows activation behavior and is highly non-linear.

- $\gamma, e|V_b|, |\epsilon_1| \ll k_B T$: High-temperature regime.

$$I_L \approx -\frac{e}{\hbar}\frac{\gamma_R \gamma_L}{\gamma_R + \gamma_L}\left(\frac{eV_b}{2k_B T}\right)\int d\varepsilon' D(\varepsilon')\frac{1}{2} \qquad (90)$$

$$\approx -\frac{e}{\hbar}\frac{\gamma_R \gamma_L}{\gamma_R + \gamma_L}\left(\frac{eV_b}{4k_B T}\right). \qquad (91)$$

The current decreases with the inverse of the temperature (linear bias regime).

- $k_B T \ll \gamma$: Low-temperature regime; strong broadening.

$$I_L \approx -\frac{e}{\hbar}\frac{\gamma_R \gamma_L}{\gamma_R + \gamma_L}\int_{-\varepsilon_1 - eV_b/2}^{-\varepsilon_1 + eV_b/2} d\varepsilon' D(\varepsilon') \qquad (92)$$

$$\approx -\frac{2e}{h}\frac{\gamma_R \gamma_L}{\gamma_R + \gamma_L}\left[\arctan\left(\frac{-\varepsilon_1 + eV_b/2}{\gamma/2}\right) - \arctan\left(\frac{-\varepsilon_1 - eV_b/2}{\gamma/2}\right)\right]. \qquad (93)$$

Notice that Eq. (93) is the starting point of a well-known theoretical description of electronic transport in "soft" molecular electronics [10,11]. The current is temperature independent and becomes linear with the bias voltage when $e|V_b| \ll \gamma$:

$$I_L \approx -\frac{e^2}{h}\frac{4\gamma_R \gamma_L}{\gamma(\gamma_R + \gamma_L)}V_b. \qquad (94)$$

# 8 Conclusions

Given that for single-channel, single-level conductance both fully coherent and sequentially incoheren approaches lead to the same expression for the current, we can conclude that the most general expression (at low bias) is given by

$$I_L = -\frac{e}{h}\frac{\gamma_R \gamma_L}{\gamma_R + \gamma_L}\int d\varepsilon \frac{\gamma}{(\varepsilon - \varepsilon_1 + eV_g)^2 + (\gamma/2)^2}\left[f\left(\frac{\varepsilon - \mu_L}{k_B T}\right) - f\left(\frac{\varepsilon - \mu_R}{k_B T}\right)\right], \qquad (95)$$

where we allow the total level width to include some broadening due to energy relaxation mechanisms other than leakage through the leads, namely, $\gamma = \gamma_R + \gamma_L + \gamma_0$.